\documentclass[aps,prd,preprint,superscriptaddress,amsmath,amssymb,nofootinbib,longbibliography,titlepage]{revtex4-2}
\usepackage[utf8]{inputenc}
\usepackage{graphicx}
\usepackage[T1]{fontenc}
\usepackage{lmodern}
\usepackage[english]{babel}
\usepackage{amsmath,amssymb}
\usepackage{xcolor}
\usepackage[colorlinks=true,linkcolor=blue,citecolor=blue,urlcolor=blue]{hyperref}
\newcommand{\x}{{\mathbf x}_\shortparallel}
\newcommand{\X}{{X}_\shortparallel}
\newcommand{\kk}{{\mathbf k}_\shortparallel}
\newcommand{\dsp}[1]{d^{#1}x_\shortparallel}
\newcommand{\dwv}{d^{\Dpar}X_\shortparallel}
\newcommand{\grad}{\nabla_\shortparallel}
\newcommand{\tr}{\operatorname{tr}}
\newcommand{\half}{\tfrac12}
\newcommand{\Dpar}{D_\shortparallel}

\begin{document}

\title{Neumann zero modes and the locality of the derivative expansion}

\author{C.~D.~Fosco}
\email{fosco@protonmail.ch}
\author{F.~D.~Mazzitelli}
\email{francisco.mazzitelli@ib.edu.ar}
\affiliation{Centro At\'omico Bariloche and Instituto Balseiro,\\
Comisi\'on Nacional de Energ\'ia At\'omica,\\
R8402AGP Bariloche, Argentina}

\begin{abstract}
The derivative expansion (DE) provides a useful approximation to the Casimir
interaction energy between two close surfaces. It has the form of a local 
functional, obtained from a zero momentum 
expansion.
However, when the system involves a massless field satisfying Neumann 
conditions, that expansion may develop nonanalytic
terms. We present here a study on how this property reflects itself in
the need to add to the DE some novel, nonlocal terms.
These are, we argue,  akin to the ones introduced in effective 
theories because of the presence of massless particles in loops.
We pinpoint the mode responsible for this effect, and construct the effective
action due to its
fluctuations, showing that it reproduces known nonanalytic contributions, 
and present the corresponding nonlocal addition to the DE.
We also predict a new nonanalyticity: Neumann conditions in $3+1$ dimensions 
at zero temperature produce a term of order $|\kk|^3$, with a parameter-free
coefficient, between the (local) second and fourth orders of the expansion.
We show that, for a cylinder in front of a plane, the zero-mode loop adds to
the local expansion a term of relative order $(L/R)^{3/2}$, with a coefficient
that we compute explicitly, showing that it accounts for the behaviour of the
existing numerical fits for that geometry.
\end{abstract}

\maketitle

\section{Introduction}
The proximity force approximation (PFA), and its systematic improvement by the derivative
expansion (DE), reduce the calculation of the Casimir interaction between two close, gently
curved surfaces to that of constructing a local functional of the distance between them
\cite{FLM2011,FLM2012,BEJK,FLM2014,review}. Locality holds whenever the form factors of
the interaction are analytic at small momenta; for a massless field with Neumann boundary
conditions, in some cases, they are not. This work is about the origin, the systematics and
the consequences of that nonanalyticity, which we interpret in the language of an effective
field theory: rather than a breakdown of the DE, it is the signature of a 
massless degree of freedom that a complete low energy description has to retain.
Note that the relevance of those terms has long been recognized in the case
of effective
theories~\cite{Donoghue1994a,Donoghue1994b,DonoghueElMenoufi}.

We consider here a plane  mirror at $x_d=0$ and a gently curved surface given 
as a Monge patch
$x_d=\psi(\x)$ over it, with $\psi>0$. This is the geometry in which the 
perturbative form factors of the DE are usually computed, and it is the one we 
shall use throughout; the general
case of two curved surfaces is discussed in \cite{BEJK}.
The analysis presented here focuses on the role of zero modes, which can already be captured in the simpler setting of a single curved surface. Before going further we fix the notation: the 
curved profile is assumed to be static, while the field lives also in time, so 
we keep two kinds of indices apart: points on the reference plane are denoted 
by $\x=(x_1,\dots,x_{d-1})$, with $d$ the number of spatial dimensions; Latin 
indices $i,j$ run over those directions, $\grad$ denotes the corresponding 
gradient, and $\psi$ depends on $\x$ alone. 
The field depends also on the Euclidean time $x_0$, and we write $\X=(x_0,\x)$ 
for the space-time coordinates that label ``parallel'' points: the mirror's 
world-volume. 
They carry  Greek indices $\mu,\nu$, $\partial_\mu=\partial/\partial
X_\shortparallel^\mu$, and we denote by $\partial^2$ the corresponding Laplacian.
$\Dpar$ denotes the number of parallel dimensions: $\Dpar=d$ at zero
temperature. When considering the sector of the vanishing Matsubara 
frequency at finite temperature the time direction is absent, and in that case
$\Dpar=d-1$.
Acting on the static profile, $\partial^2=\grad^2$. 
Capital indices $M,N$ run over the $\Dpar+1$
coordinates $(\X,u)$ of the region between the mirrors, $u$ being a
 coordinate introduced in Section \ref{sec:flat}. Finally, $\kk$ is
the momentum conjugate to $\x$, while $p$ and $l$ denote momenta conjugate to
$\X$.

In this notation, the DE organizes the interaction energy of a massless field as a local
functional of the profile \cite{FLM2011,FLM2012,BEJK,FLM2014,review},
\begin{equation}
F=\int \dsp{d-1}\;\frac{1}{\psi(\x)^{d}}\Big[\,b_0+b_2\,\big|\grad\psi\big|^{2}
+O(\partial^4)\,\Big]\,,
\label{eq:DE}
\end{equation}
$b_0$ being the parallel-plate coefficient, which defines the PFA,
and $b_2$ the next-to-leading order coefficient. These coefficients are obtained
from the perturbative form factors of the specific interaction. For Dirichlet 
conditions the coefficients are finite, and the expansion is local order by 
order.

For Neumann conditions this is not always the case: in $d=2$ at zero 
temperature, and in $d=3$
at any positive temperature, the next-to-leading order form factor is nonanalytic at zero momentum, so that the
second term in \eqref{eq:DE} becomes nonlocal in coordinate space. This is 
indeed pointed out in  \cite{FLMdT}, and  restated in the review \cite{review}, where the
small momentum expansion of the form factor is quoted as containing, besides the analytic
$|\kk|^2$, a term proportional to $(Ta)\,|\kk|^2\log(|\kk|^2a^2)$. The cause is identified in
\cite{FLMdT,review} as the existence of a gapless excitation, noted in \cite{FLMdT} to be
present also for $d>2$, where it produces subleading terms; we show below what those terms
are for $d=3$ at zero temperature, and compute their coefficient. The same
effect is seen in
\cite{FLMimperfect}, where imperfect mirrors in $2+1$ dimensions develop the 
nonanalyticity
only in the limit of perfect conductivity, the imperfection acting as an infrared regulator,
and in \cite{FLMhighT}, where the nonanalyticity of the electromagnetic free energy at high
temperature is traced to the dimensionally reduced thermal zero mode.

This situation is a familiar one in effective field theories, and it is useful 
to state it in that language from the outset: integrating out
heavy degrees of freedom produces a local effective action, an expansion in derivatives over
the lightest of the masses involved. Massless particles cannot be treated in this way: they
have to be kept as explicit fields of the effective theory, and their loops generate
contributions which are nonanalytic in the momenta, such as $\log(-q^2)$ or $\sqrt{-q^2}$,
and therefore nonlocal in coordinate space. Donoghue's treatment of general relativity as an
effective field theory \cite{Donoghue1994a,Donoghue1994b} is the standard example. There, the
nonanalytic terms generated by loops of massless gravitons are cleanly separated from the
local ones. The latter produce contact terms for the potential between two 
masses, while the nonanalytic ones  dominate at large distances, and their 
coefficients do not depend on the unknown ultraviolet
completion: they are parameter free predictions of the low energy theory. The 
same terms can be organized in a covariant nonlocal action, as a curvature 
expansion with nonlocal form factors \cite{DonoghueElMenoufi}.

The Casimir problem has exactly this structure. The interaction energy is the 
effective action for the geometry, obtained by integrating out the field. In 
the gap between the mirrors the field decomposes into a tower of  
modes which, seen from the parallel
directions, are massive, with masses of order $1/\psi$; integrating them out yields a local
functional, an expansion in $\psi\partial$, and that functional is the DE. For Neumann
conditions, however, the lowest mode of the tower is massless, and a symmetry keeps it so for
any profile. This mode plays the role of the graviton. An appropriate effective 
description must keep it as an explicit field, with local couplings to the 
geometry that the DE itself determines, and its loop produces nonlocal terms in 
the interaction energy. 
From this point of view the nonanalytic terms are not a failure of the 
expansion but its most robust output: their coefficients follow from the low 
energy sector alone, and they appear at orders that no local term can reach. We 
shall find two instances: a term of order $|\kk|^3$ in the Neumann
form factor in $3+1$ dimensions, and a half integer power of $L/R$ in the 
corrections to the PFA for a cylinder of radius $R$ at a distance $L$ from a 
plane.

The structure of this paper is as follows: in Section~\ref{sec:aux} we identify 
the massless mode in the auxiliary field  language of
\cite{FLM2011,FLM2012,FLMdT,review}, where it shows up as a zero of one of the 
eigenvalues of the kernel $\mathbb U$, and derive the rule from the infrared 
behaviour of a  one loop integral. Section~\ref{sec:flat} identifies the zero 
mode in the context of a flattening choice of coordinates, and a Kaluza-Klein 
reduction. In Section~\ref{sec:effective} we present the effective action 
calculation, and Section~\ref{sec:examples} applies our results to the concrete 
case of a cylinder in front of a plane. Section~\ref{sec:discussion} assembles 
the resulting procedure and discusses its scope. Tests against previous results
are collected in Appendix~\ref{app:check}, and the numerical evaluation of the
cylinder-plane coefficient is described in Appendix~\ref{app:kappa}.

\section{The zero mode in the auxiliary field formulation}
\label{sec:aux}
The mechanism, and its main quantitative consequence, can be obtained entirely 
within the auxiliary field representation of the Casimir energy used in
\cite{FLM2011,FLM2012,FLMdT,review}. We do that first: it is the shortest route to the answer,
and it makes contact with the form factors as they are actually computed.

In that representation the Neumann conditions on the two mirrors are imposed by delta
functionals, written in terms of auxiliary fields $\lambda_1,\lambda_2$ living on them. The field is then
Gaussian and can be integrated out, leaving \cite{FLMdT}
\begin{equation}
\mathcal Z_\beta(\psi)=\mathcal Z^{(0)}_\beta\int\mathcal D\lambda_1\,\mathcal D\lambda_2\;
e^{-\frac12\int\int\lambda_A\,\mathbb U_{AB}\,\lambda_B}\,,
\qquad
\Gamma_\beta(\psi)=\frac{1}{2\beta}\,\tr\log\mathbb U\,,
\label{eq:Ukernel}
\end{equation}
with $\mathbb U_{AB}$ built from normal derivatives of the free propagator evaluated on the two
surfaces, and $\Gamma_\beta$ the free energy. One feature of that construction returns below in
geometric guise: the area element $\sqrt g$ accompanying the delta functional on the curved
mirror is absorbed by a rescaling of $\lambda_2$, since $\sqrt g\,\partial_n=\partial_N$ with
$\partial_N\equiv\partial_d-\partial_i\psi\,\partial_i$. The geometric measure never appears
on its own.

The massless mode is already visible for parallel plates. For $\psi=a$ constant, in momentum
space for the parallel coordinates,
\begin{equation}
\mathbb U^{\rm N}=-\frac{|p|}{2}
\begin{pmatrix}1&e^{-|p|a}\\ e^{-|p|a}&1\end{pmatrix}\,,
\qquad
\mathbb U^{\rm D}=\frac{1}{2|p|}
\begin{pmatrix}1&e^{-|p|a}\\ e^{-|p|a}&1\end{pmatrix}\,,
\label{eq:Uflat}
\end{equation}
so that $\mathbb U^{\rm N}=-p^{2}\,\mathbb U^{\rm D}$: the two kernels differ by one normal
derivative per surface, an ultralocal factor, which is why both give the same interaction
energy, $\tfrac{1}{2\beta}\tr\log(1-e^{-2|p|a})$.

Their infrared behaviour is not the same. Both are diagonalized by the combinations
$\lambda_\pm=(\lambda_2\pm\lambda_1)/\sqrt2$,
\begin{equation}
\mathbb U^{\rm N}_{\pm\pm}=-\frac{|p|}{2}\big(1\pm e^{-|p|a}\big)\,,
\qquad
\mathbb U^{\rm D}_{\pm\pm}=\frac{1}{2|p|}\big(1\pm e^{-|p|a}\big)\,,
\end{equation}
and the eigenvalue corresponding to the antisymmetric combination behaves, at small momenta, as
\begin{equation}
\mathbb U^{\rm N}_{--}\;\longrightarrow\;-\frac{a}{2}\,p^{2}\,,
\qquad
\mathbb U^{\rm D}_{--}\;\longrightarrow\;\frac{a}{2}\,.
\label{eq:channels}
\end{equation}
The factor $1-e^{-|p|a}\to|p|a$ is common to both; what differs is whether the prefactor
compensates it. For Dirichlet it does: $\mathbb U^{\rm D}_{--}$ tends to a constant, so that
combination does not propagate and $[\mathbb U^{\rm D}]^{-1}$ is regular at $p=0$. For Neumann
it reinforces it, and $\mathbb U^{\rm N}_{--}$ vanishes quadratically: this is the inverse
propagator of a massless field which propagates along the mirrors, with a coefficient of $p^2$
proportional to the distance between them. It is the same field as the $\phi_0$ of \eqref{eq:S0} below, obtained there by a different route. The symmetric combinations tend to $-|p|$ and $1/|p|$, twice the kernels of a single isolated mirror, and have no such zero.

Its consequence for the form factor can be read off the one loop expression obtained by
expanding $\tr\log\mathbb U[\psi]$ about $\psi=a+\eta(\x)$. In the normalization of
\cite{FLMdT},
\begin{equation}
\Gamma^{(2)}=\frac12\int\frac{d^{d-1}k_\shortparallel}{(2\pi)^{d-1}}\;g^{(2)}(\kk)\,
\big|\tilde\eta(\kk)\big|^{2}\,,
\qquad
g^{(2)}=a^{-(d+2)}\,G^{(2)}\big(d,a|\kk|\big)\,,
\label{eq:ffnorm}
\end{equation}
and at zero temperature, in units $a=1$, the Neumann and Dirichlet form factors are
\begin{equation}
G^{(2)}_{\rm N}(d,l)=-2\!\int\!\frac{d^{d}p}{(2\pi)^{d}}
\frac{\big[p\cdot(p+l)\big]^{2}}{|p|\,|p+l|}\;\mathcal K\,,
\qquad
G^{(2)}_{\rm D}(d,l)=-2\!\int\!\frac{d^{d}p}{(2\pi)^{d}}\;|p|\,|p+l|\;\mathcal K\,,
\label{eq:twoff}
\end{equation}
with the same loop denominators in both,
$\mathcal K=\big[1-e^{-2|p|}\big]^{-1}\big[e^{2|p+l|}-1\big]^{-1}$. These are the expressions
obtained in \cite{FLMdT}. The two internal lines carry a simple pole at zero momentum,
$\mathcal K\to1/4|p||p+l|$, and that pole is the near zero eigenvalue of 
\eqref{eq:channels}. Since the rest of the integrand is
regular at $p=0$ and at $p=-l$, this is the only region that can generate a nonanalyticity in
$l$. Keeping it,
\begin{equation}
G^{(2)}_{\rm D}\Big|_{\rm IR}=-\frac12\int\frac{d^{d}p}{(2\pi)^{d}}\,1\,,
\qquad
G^{(2)}_{\rm N}\Big|_{\rm IR}=-\frac12\int\frac{d^{d}p}{(2\pi)^{d}}\;
\frac{\big[p\cdot(p+l)\big]^{2}}{p^{2}\,(p+l)^{2}}\,.
\label{eq:IRtwo}
\end{equation}
The contrast is clear. The Dirichlet numerator cancels the two poles identically and leaves an
integral with no dependence on $l$: no nonanalyticity is generated. The Neumann numerator
supplies one further inverse power of each momentum, and the two poles become two massless
propagators. Writing $p\cdot(p+l)=\half\big[p^{2}+(p+l)^{2}-l^{2}\big]$ and expanding the
square,
\begin{equation}
\frac{\big[p\cdot(p+l)\big]^{2}}{p^{2}(p+l)^{2}}
=\frac14\Big[\frac{p^{2}}{(p+l)^{2}}+\frac{(p+l)^{2}}{p^{2}}+2
-\frac{2l^{2}}{p^{2}}-\frac{2l^{2}}{(p+l)^{2}}+\frac{l^{4}}{p^{2}(p+l)^{2}}\Big]\,,
\label{eq:decomp}
\end{equation}
every term is analytic in $l^{2}$ upon integration, by power counting and a shift of the
integration variable, except the last, which is the massless bubble. Hence
\begin{equation}
\begin{gathered}
G^{(2)}_{\rm N}\Big|_{\rm nonanalytic}=-\frac{l^{4}}{8}\,B_{\Dpar}(l)\,,
\\[4pt]
B_{D}(l)=\int\frac{d^{D}q}{(2\pi)^{D}}\frac{1}{q^{2}(q+l)^{2}}
=\frac{\Gamma\big(2-\tfrac D2\big)\Gamma\big(\tfrac D2-1\big)^{2}}
{(4\pi)^{D/2}\,\Gamma(D-2)}\;\big(l^{2}\big)^{\frac D2-2}\,,
\end{gathered}
\label{eq:predff}
\end{equation}
evaluated at $D=\Dpar=d$ here. In dimensional regularization this gives, for $\Dpar\neq2$,
\begin{equation}
G^{(2)}_{\rm N}\Big|_{\rm nonanalytic}=C_{\Dpar}\,|l|^{\Dpar}\,,
\qquad
C_{D}=-\frac{\Gamma\big(2-\tfrac D2\big)\Gamma\big(\tfrac D2-1\big)^{2}}
{8\,(4\pi)^{D/2}\,\Gamma(D-2)}\,,
\label{eq:CD}
\end{equation}
while $\Dpar=2$ is the logarithmic case. Three values are used below: $C_{1}=0$, since
$1/\Gamma(D-2)$ vanishes at $D=1$; $C_{3}=-1/64$; and for $\Dpar=2$, $-l^{2}\log l^{2}/16\pi$,
which is the result of \cite{FLMdT}. The argument just given is the one used there for that
particular case, and nothing in it is special to $d=2$.

Equation \eqref{eq:predff} already contains the main result, but it leaves three questions
open. Why is that eigenvalue gapless, and can a deformation of the mirror lift it? Section
\ref{sec:KK} shows that a symmetry protects it. What replaces \eqref{eq:channels} when
$\psi$ is not a small deformation of a constant? A Kaluza-Klein reduction on the gap, carried
out in Sections \ref{sec:flat}, \ref{sec:KK} and \ref{sec:effective}. And what functional of the
geometry resums the effect beyond second order in the deformation? A massless determinant in a
local potential built from $\psi$, obtained in Section \ref{sec:exact}.

\section{Flattening the gap}
\label{sec:flat}

A convenient way to identify the origin of the massless mode is to flatten the gap by a change
of coordinates. We replace $(\X,x_d)$ by $(\X,u)$, with
\begin{equation}
x_d\;=\;u\,\psi(\x)\,,
\qquad\text{equivalently}\qquad
u=\frac{x_d}{\psi(\x)}\in[0,1]\,.
\label{eq:flatten}
\end{equation}
The plane is now $u=0$ and the curved mirror is $u=1$, for every $\x$. The coordinate $u$ is
dimensionless, and the geometry has been transferred from the domain to the metric.
Since $\X$ is unchanged and $dx_d=A_\mu dX_\shortparallel^\mu+\psi\,du$
with
\begin{equation}
A_\mu\;\equiv\;\partial_\mu\big[u\,\psi\big]\,,\qquad
A_0=0\,,\qquad A_i=u\,\partial_i\psi\,,
\label{eq:Adef}
\end{equation}
the flat Euclidean line element
$ds^2=\delta_{\mu\nu}dX_\shortparallel^\mu dX_\shortparallel^\nu+dx_d^2$ becomes
\begin{equation}
ds^2=\big(\delta_{\mu\nu}+A_\mu A_\nu\big)\,dX_\shortparallel^\mu dX_\shortparallel^\nu
\;+\;2\,\psi A_\mu\,dX_\shortparallel^\mu du\;+\;\psi^2\,du^2\,,
\end{equation}
that is
\begin{equation}
g_{\mu\nu}=\delta_{\mu\nu}+A_\mu A_\nu\,,\qquad
g_{\mu u}=\psi\,A_\mu\,,\qquad
g_{uu}=\psi^2\,.
\label{eq:metric}
\end{equation}
The shear vector $A_\mu$ has no time component, because the profile is static; it grows
linearly from zero on the plane to $\grad\psi$ on the curved mirror. The inverse metric is
\begin{equation}
g^{\mu\nu}=\delta^{\mu\nu}\,,\qquad
g^{\mu u}=-\frac{A^\mu}{\psi}\,,\qquad
g^{uu}=\frac{1+A^2}{\psi^2}\,,\qquad A^2\equiv A_\mu A_\mu=u^2|\grad\psi|^2\,,
\label{eq:inversemetric}
\end{equation}
as follows by direct multiplication, and the determinant is
\begin{equation}
\det g\;=\;\psi^2\,,\qquad\text{hence}\qquad \sqrt{g}\;=\;\psi(\x)\,,
\label{eq:detg}
\end{equation}
which is immediate from \eqref{eq:flatten}: the change of coordinates is a shear in the $x_d$
direction with Jacobian $\partial x_d/\partial u=\psi$. Note that $\sqrt g$ does not depend on
$u$. The frame dual to \eqref{eq:metric} gives the familiar mixing of derivatives,
\begin{equation}
\partial_\mu\big|_{x_d}=\partial_\mu\big|_{u}-\frac{A_\mu}{\psi}\,\partial_u\,,
\qquad
\partial_d=\frac{1}{\psi}\,\partial_u\,,
\label{eq:derivmix}
\end{equation}
so that the Euclidean action of a massless scalar in the region between the mirrors is
\begin{equation}
S=\half\int \dwv du\;\sqrt g\;g^{MN}\partial_M\phi\,\partial_N\phi
=\half\int \dwv du\;\psi(\x)\,
\Big[\Big(\partial_\mu\phi-\frac{A_\mu}{\psi}\partial_u\phi\Big)^2
+\frac{1}{\psi^2}\big(\partial_u\phi\big)^2\Big]\,.
\label{eq:flataction}
\end{equation}
The whole deviation from the block diagonal form $g_{MN}=\mathrm{diag}(\delta_{\mu\nu},\psi^2)$
is carried by $A_\mu$, of first order in $\grad\psi$. Equations \eqref{eq:metric} to
\eqref{eq:flataction} involve no approximation.

The boundary conditions take a simple form in these coordinates. Since the two mirrors are
at constant values of  $u$, the unit conormal is $n_M=\pm\,\partial_M u/\sqrt{g^{uu}}$, and the normal derivative is
\begin{equation}
n^M\partial_M\phi=\pm\frac{1}{\sqrt{1+A^2}}
\Big[\frac{1+A^2}{\psi}\,\partial_u\phi-A^i\partial_i\phi\Big]\,,
\label{eq:normalderiv}
\end{equation}
so that the Neumann condition $n^M\partial_M\phi=0$ reads, with no approximation,
\begin{equation}
\partial_u\phi\Big|_{u=0}=0\,,
\qquad
\partial_u\phi\Big|_{u=1}=\frac{\psi\;\grad\psi\!\cdot\!\grad\phi}{1+|\grad\psi|^{2}}\,.
\label{eq:neumannflat}
\end{equation}
On the plane the condition is exactly the flat one, because $A_i$ vanishes there. On the curved
mirror it is inhomogeneous, of first order in $\grad\psi$, and the Monge area element
$\sqrt{1+|\grad\psi|^2}$ has dropped out of the homogeneous condition. Dirichlet conditions are
$\phi=0$ at $u=0,1$ and are unaffected by the change of coordinates.

The inhomogeneous term in \eqref{eq:neumannflat} does all the work in Section \ref{sec:effective}:
because of it a $u$ independent field cannot satisfy the boundary conditions once it varies
along the mirrors. Note also that \eqref{eq:neumannflat} is the natural boundary condition of
the variational problem defined by \eqref{eq:flataction}: varying $S$ with no condition imposed
on $\phi$ at $u=0,1$ produces the boundary term
$\oint\sqrt g\,g^{uN}\partial_N\phi\,\delta\phi$, whose vanishing is \eqref{eq:neumannflat}.

Finally, the measure in field space is the one induced by the metric,
\begin{equation}
\|\delta\phi\|^2=\int \dwv du\;\sqrt g\,(\delta\phi)^2
=\int \dwv du\;\psi(\x)\,(\delta\phi)^2\,,
\label{eq:measure}
\end{equation}
a fact used in Section \ref{sec:exact}, where the factor $\psi$ cancels a contact term.

\subsection{The gap as a Kaluza-Klein interval}
\label{sec:KK}

At zeroth order in derivatives, that is setting $A_\mu\to0$ in \eqref{eq:flataction} and
\eqref{eq:neumannflat}, the action is
\begin{equation}
S^{(0)}=\half\int \dwv du\;\psi(\x)
\Big[(\partial_\mu\phi)^2+\frac{1}{\psi(\x)^2}(\partial_u\phi)^2\Big]\,,
\label{eq:S0flat}
\end{equation}
and the transverse problem is the interval $u\in[0,1]$ with the appropriate conditions at its
ends. The gain of the flattening is that this interval is the same for every $\x$, so a single
$\x$-independent orthonormal basis of eigenfunctions of $-\partial_u^2$ diagonalizes it
globally. For Neumann conditions at both ends,
\begin{equation}
f_n(u)=\sqrt{2-\delta_{n0}}\,\cos(n\pi u)\,,\qquad \lambda_n=n\pi\,,\qquad n\ge0\,,
\label{eq:basis}
\end{equation}
with $f_0=1$. Dirichlet at both ends would give $\sqrt2\sin(n\pi u)$ with $n\ge1$, and
mixed boundary conditions give $\sqrt2\sin[(n+\half)\pi u]$ with $n\ge0$: in the last  two cases the lowest eigenvalue
is $\pi$ and $\pi/2$ respectively, while the Neumann tower starts at $\lambda_0=0$.

Expanding the field in this basis, $\phi(\X,u)=\sum_n f_n(u)\,\phi_n(\X)$, and inserting it
in \eqref{eq:S0flat}, the $u$ integration is done once and for all, and
\begin{equation}
S^{(0)}=\half\int \dwv\;\psi(\x)\sum_n
\Big[(\partial_\mu\phi_n)^2+m_n(\x)^2\,\phi_n^2\Big]\,,
\qquad
m_n(\x)=\frac{n\pi}{\psi(\x)}\,.
\label{eq:KKaction0}
\end{equation}
This is a Kaluza-Klein reduction with $\x$ dependent masses, and it is worth being explicit
about what that means, since the original field is massless and no mass has been added by hand.
The mass of $\phi_n$ is the eigenvalue of the transverse Laplacian, which in the flattened
chart is $-\psi(\x)^{-2}\partial_u^2$: the geometry has been moved out of the domain and into
the coefficient multiplying $\partial_u^2$. Since the $f_n$ are fixed functions of $u$ alone,
the entire $\x$ dependence of the transverse spectrum resides in that prefactor, and the
eigenvalues $n^2\pi^2/\psi(\x)^2$ act, from the point of view of the parallel directions,
as a position dependent mass squared. It is the local cutoff frequency of a waveguide of width
$\psi(\x)$: a mode of the tower propagates along the mirrors only where it is above its local
threshold. The mass depends on $\x$ and not on $x_0$: the
transverse spectrum is static, as it must be for static mirrors.

Two remarks on the status of \eqref{eq:KKaction0}. First, $m_n(\x)^2$ multiplies $\phi_n^2$ and
not a derivative, so no operator ordering ambiguity arises in it; the ambiguities live in the
terms discarded when setting $A_\mu\to0$. Second, treating the local transverse spectrum as if
it were the exact one is an adiabatic approximation, controlled by $|\grad\psi|\ll1$, which is
the regime the DE assumes. The corrections are the mixing terms generated by $A_i$, and they
are of higher order in derivatives; Section \ref{sec:effective} computes the first of them.

This makes the structure of the DE transparent. Integrating out a mode of mass $m_n$ produces
form factors whose nearest singularity in the transverse momentum sits at a momentum of order
$m_n$: the two particle threshold $|p|=2m_n$ for the determinant of the tower, and $|p|=m_n$
for the couplings it induces on the zero mode. Their low momentum expansion is therefore an
expansion in
$\partial/m_n\sim\psi\partial$, which is the DE parameter, and it is local. What controls it is
the infimum of $m_1$ over the base, $\pi/\sup_{\x}\psi$, which is nonzero for any configuration
of bounded gap. A gapless mode has no such threshold: the branch point sits at the origin.

The criterion is therefore the following: the source of nonlocality studied here is present
precisely when the Kaluza-Klein spectrum  is gapless, the expansion parameter being
$\psi\partial$, that is, the ratio of the derivative scale to the lowest Kaluza-Klein mass.
Dirichlet and mixed conditions are gapped and no nonanalyticity of this kind arises from them.
Neumann is not gapped. The dimension and the temperature only control the
derivative order at which the massless mode shows up, and in what form, as Section
\ref{sec:where} shows. The criterion concerns this mechanism
only; it does not exclude nonlocalities of a different origin, such as those associated with
the finite lateral size of the mirrors.

The gaplessness of the $n=0$ Neumann mode is not an artifact of the leading order. For a
massless field with Neumann conditions on all boundaries,
\begin{equation}
\phi\;\longrightarrow\;\phi+c
\label{eq:shift}
\end{equation}
is a symmetry of the action and of the boundary conditions for an arbitrary profile, and
$\phi=\mathrm{const}$ solves the equation of motion. This holds for the unapproximated
condition \eqref{eq:neumannflat}, not only for its zeroth order form. Dirichlet conditions
break \eqref{eq:shift} explicitly, and so do a mass term and a Robin condition.

Two consequences follow. First, no mass term for $\phi_0$ can be generated at any order in the
derivative expansion of the mixing terms, since $f_0=1$ and therefore $\phi_0\to\phi_0+c$ under
\eqref{eq:shift}: the zero mode stays massless however curved the mirror is. Second, the
nonlocal terms disappear under any deformation that breaks the shift symmetry, which is the
 reason why imperfect mirrors, dissipation and mixed Dirichlet-Neumann boundary conditions admit a
purely local expansion.

\section{The effective action for the zero mode}
\label{sec:effective}

We now derive the action that governs $\phi_0$, by integrating out the massive modes
$\phi_{n\ge1}$. Since the action is quadratic, the integral is Gaussian and the derivation is
elementary; it is also the point where the geometry of the mirror enters the zero mode sector.

Consider first a configuration with no $u$ dependence, $\phi=\phi_0(\X)$. Every term
containing $A_i$ in \eqref{eq:flataction} drops out, since all of them carry
$\partial_u\phi$, and
\begin{equation}
S[\phi_0]=\half\int \dwv\;\psi(\x)\,\big(\partial_\mu\phi_0\big)^2\,.
\label{eq:S0}
\end{equation}
This is the action of a massless field in the parallel directions, with the local width of the
gap as a position dependent coefficient. It cannot be the whole answer: a $u$ independent
configuration violates the boundary condition on the curved mirror in \eqref{eq:neumannflat}
or, in the language of the mode expansion, it neglects the coupling between $\phi_0$ and the
massive modes, to which we now turn.

To include that coupling, we insert
\begin{equation}
\phi(\X,u)=\phi_0(\X)+\sum_{n\ge1}\phi_n(\X)\,f_n(u)\,,
\qquad f_n(u)=\sqrt2\,\cos(n\pi u)\,,
\label{eq:split}
\end{equation}
into \eqref{eq:flataction}. There is no mixing between $\partial_\mu\phi_0$ and
$\partial_\mu\phi_n$, since $\int_0^1du\,f_n=0$ for every $n\ge1$; the only mixing comes from
the term linear in $A_i$ produced by expanding the square in \eqref{eq:flataction}, namely
$-A_i\,\partial_i\phi_0\,\partial_u\phi$, and it involves the overlaps
\begin{equation}
\int_0^1du\;A_i\,\partial_uf_n=\partial_i\psi\int_0^1du\;u\,\partial_uf_n
=\sqrt2\,(-1)^{n}\,\partial_i\psi\,,
\label{eq:overlap}
\end{equation}
where we used $A_i=u\,\partial_i\psi$. Therefore
\begin{equation}
\begin{gathered}
S=\half\int\dwv\;\psi\big(\partial_\mu\phi_0\big)^2
+\int\dwv\sum_{n\ge1}\Big[\frac{n^2\pi^2}{2\psi}\,\phi_n^2-J_n\,\phi_n\Big]
\;+\;\dots\,,
\\[4pt]
J_n=\sqrt2\,(-1)^{n}\,\grad\psi\!\cdot\!\grad\phi_0\,.
\end{gathered}
\label{eq:Squad}
\end{equation}
Every massive mode is coupled to one and the same source, the projection
$\grad\psi\cdot\grad\phi_0$ of the field gradient on the slope of the curved mirror, with
alternating sign and equal strength. The dots stand for two kinds of terms: those in which the
massive modes appear with parallel derivatives, suppressed by powers of $\partial/m_n$
relative to the ones kept, and those quadratic in $A_i$, of higher order in the slope.

The integral over each $\phi_n$ in \eqref{eq:Squad} is then Gaussian, with a term linear in
the integration variable. Completing the
square,
\begin{equation}
\frac{n^2\pi^2}{2\psi}\,\phi_n^2-J_n\,\phi_n
=\frac{n^2\pi^2}{2\psi}\,\big(\phi_n-\bar\phi_n\big)^2
-\frac{\psi\,J_n^2}{2\,n^2\pi^2}\,,
\qquad
\bar\phi_n=\frac{\psi\,J_n}{n^2\pi^2}\,,
\label{eq:square}
\end{equation}
the integral over the shifted variable produces a determinant which does not involve $\phi_0$;
restoring the parallel derivatives of $\phi_n$, that determinant becomes the contribution of
the gapped modes to the interaction energy, which is local by the argument of Section
\ref{sec:KK} and belongs to the ordinary DE. What survives in the effective action for
$\phi_0$ is the last term of \eqref{eq:square}, summed over the tower:
\begin{equation}
\Delta S=-\int\dwv\sum_{n\ge1}\frac{\psi\,J_n^2}{2\,n^2\pi^2}
=-\frac16\int\dwv\;\psi\,\big(\grad\psi\!\cdot\!\grad\phi_0\big)^2\,,
\label{eq:DeltaS}
\end{equation}
where we used $J_n^2=2\,(\grad\psi\!\cdot\!\grad\phi_0)^2$ and
$\sum_{n\ge1}n^{-2}=\zeta(2)=\pi^2/6$. Adding \eqref{eq:DeltaS} to \eqref{eq:S0},
\begin{equation}
S_0^{\rm eff}[\phi_0]=\half\int\dwv\;Z_{\mu\nu}\,\partial_\mu\phi_0\,\partial_\nu\phi_0\,,
\qquad
\begin{aligned}
Z_{00}&=\psi\,,\qquad Z_{0i}=0\,,\\
Z_{ij}&=\psi\Big[\delta_{ij}-\tfrac13\,\partial_i\psi\,\partial_j\psi\Big]+O(\partial^4)\,.
\end{aligned}
\label{eq:Z}
\end{equation}
The time direction is untouched, as it must be for static mirrors, and the correction is a
projector along the slope: the zero mode propagates with a smaller coefficient in the
direction of $\grad\psi$ than in the directions orthogonal to it.

\subsection{Integration of the zero mode}
\label{sec:exact}

The quadratic operator defined by \eqref{eq:S0} is
$\mathcal O=-\partial_\mu\,\psi\,\partial_\mu$. Two facts make this sector tractable. The first
is a similarity transformation: for any smooth positive $\psi$ and in any dimension,
\begin{equation}
-\partial_\mu\,\psi\,\partial_\mu\;=\;\sqrt \psi\,\Big(-\partial^2+V\Big)\,\sqrt \psi\,,
\qquad V\;=\;\frac{\partial^2\sqrt \psi}{\sqrt \psi}
\;=\;\frac{\grad^2\psi}{2\psi}-\frac{|\grad\psi|^2}{4\psi^2}\,,
\label{eq:simil}
\end{equation}
as follows by acting on $\psi^{-1/2}f$: the two first derivative terms cancel. Since $\psi$ is static, $\partial^2$ here is the
spatial Laplacian, and $V$ is a geometric object of second derivative order: to leading order
$\grad^2\psi$ is the trace of the second fundamental form of the curved mirror, so $V$ is that
trace over twice the gap, corrected by a slope term. It vanishes whenever $\sqrt\psi$
is harmonic on the base, in particular for parallel plates.

Note also that \eqref{eq:simil} implies $(-\partial^2+V)\sqrt \psi=0$: the image of the
constant under the similarity transformation is an exact zero mode of $-\partial^2+V$. This is the shift symmetry \eqref{eq:shift} after the reduction, and it means that the determinant
below must be understood with that mode removed.

The second fact is that the measure cancels the contact term produced by this
transformation. By \eqref{eq:measure}, the metric on field space inherited from the
flattened action is $\|\delta\phi_0\|^2=\int \dwv\,\psi\,(\delta\phi_0)^2$, and the
Gaussian integration therefore
produces $\det'(\psi^{-1}\mathcal O)^{-1/2}$ and not $\det'(\mathcal O)^{-1/2}$, the prime
denoting omission of the zero mode. By \eqref{eq:simil},
$\psi^{-1}\mathcal O=\psi^{-1/2}(-\partial^2+V)\psi^{1/2}$, which is similar to
$-\partial^2+V$. The ultralocal factor $\tr\log \psi$ therefore cancels, with no
renormalization ambiguity, and
\begin{equation}
\Gamma_0\;=\;\half\log{\det}'\Big(-\partial^2+V\Big)\,,
\qquad V=\frac{\partial^2\sqrt \psi}{\sqrt \psi}\,.
\label{eq:Gamma0}
\end{equation}
The contribution of the gapless sector is thus the determinant of a massless field in the parallel directions, in a local potential built from the profile. This is the technical content of extracting the zero
mode. In the language of Section \ref{sec:aux}, \eqref{eq:Gamma0} is the contribution of the antisymmetric combination to $\tr\log\mathbb U$, resummed in the deformation and rewritten as a
functional of $\psi$; expanding it to second order returns \eqref{eq:predff}, as Section
\ref{sec:where} verifies.

Subtracting the parallel plate reference,
\begin{equation}
\Gamma_0=\half\tr\log\big(1+G_0V\big)\,,\qquad G_0=\big(-\partial^2\big)^{-1}\,,
\label{eq:expand}
\end{equation}
whose first order term $\half G_0(0)\int V$ is a contact term, scaleless and vanishing in
dimensional regularization, and in any case local and absorbed in the DE. The nonanalyticity
starts at second order,
\begin{equation}
\Gamma_0^{(2)}=-\frac14\int\frac{d^{\Dpar}p}{(2\pi)^{\Dpar}}\,\big|\tilde V(p)\big|^2\;
B_{\Dpar}(p)\,,
\label{eq:Gamma2}
\end{equation}
with $B_D$ the massless bubble of \eqref{eq:predff}. The omission of the strict zero mode in
$\det'$ matters only for $\Dpar\le2$, where $G_0(p)=p^{-2}$ is not integrable at the origin;
this is the source of the infrared scale discussed in Section \ref{sec:where}.

\subsection{The nonlocal terms and their size}
\label{sec:where}

With $\psi=\bar\psi(1+h)$ one has $V=\half\grad^2 h+O(h^2)$, so that
$\tilde V(p)=-\half p^2\tilde h(p)+O(h^2)$, and \eqref{eq:Gamma2} gives
\begin{equation}
\Gamma_0^{(2)}=-\frac{1}{16}\int\frac{d^{\Dpar}p}{(2\pi)^{\Dpar}}\;p^4\,B_{\Dpar}(p)\;
\big|\tilde h(p)\big|^{2}+O(h^3)\,,
\qquad
p^4B_{\Dpar}(p)\propto|p|^{\Dpar}\,,
\label{eq:rule}
\end{equation}
in agreement with \eqref{eq:predff}, as it must be. The rule is that the Neumann zero mode
adds to the derivative expansion nonlocal terms starting at derivative order $\Dpar$, the number of parallel dimensions, with the coefficient $C_{\Dpar}$ of \eqref{eq:CD}: the
nonanalyticity is logarithmic when $\Dpar$ is even, an odd power of $|p|$ when $\Dpar$ is odd,
and it is absent at $\Dpar=1$, where $C_1=0$. The even and odd alternation is the same one
found in the covariant nonlocal expansion of boundary effective actions \cite{BFGM}, where the
form factors are $(-\partial^2)^{D/2-1}\log(-\partial^2/\mu^2)$ in even and
$(-\partial^2)^{D/2-1}$ in odd boundary dimension; here it is inherited by the two mirror
problem through \eqref{eq:Gamma0}.

Explicitly, using $B_3(p)=1/8|p|$ and $B_2(p)=\log(p^2/\mu^2)/2\pi p^2$, the second with an
infrared scale $\mu$ discussed below, and reducing to a static profile so that
$\tilde h(p)=2\pi\delta(\omega)\tilde h(\kk)$, one obtains the nonanalytic pieces of the
interaction energy. In $d=3$ at zero temperature, where $\Dpar=3$, the zero mode contributes
\begin{equation}
E_{\rm na}=-\frac{1}{128\,\bar \psi^{2}}\int\frac{d^2k_\shortparallel}{(2\pi)^2}\;|\kk|^{3}\;
\big|\widetilde{\delta \psi}(\kk)\big|^{2}\,,
\label{eq:d3T0}
\end{equation}
a finite term of derivative order three: one power of $\psi|\kk|$ beyond the next-to-leading
order and one below the fourth order. In coordinate space the corresponding kernel falls off as
$|\x|^{-5}$ in the transverse plane, so the contribution is long ranged. This term has not
been identified in the literature, although, as shown in Appendix~\ref{app:check}, it is
contained in the closed form corrugation kernel of \cite{EHGK1,EHGK2}.

The case $\Dpar=2$ covers two situations, $d=2$ at zero temperature and $d=3$ at finite
temperature. In the first one
\begin{equation}
E_{\rm na}=-\frac{1}{32\pi\,\bar \psi^{2}}
\int\frac{d^{d-1}k_\shortparallel}{(2\pi)^{d-1}}\;|\kk|^{2}\,
\log\frac{|\kk|^{2}}{\mu^{2}}\;\big|\widetilde{\delta \psi}(\kk)\big|^{2}\,,
\label{eq:D2}
\end{equation}
a term which enters already at the next-to-leading order; in the second one, the same
expression multiplied by $T$ is the nonanalytic part of the free energy, for the reason
explained below. These are the two cases established
in \cite{FLMdT} and quoted in the discussion of Neumann conditions at finite
temperature in \cite{review}. The coefficient of the
logarithm is fixed by the zero mode; the scale inside it is not, and is fixed by matching to
the gapped sector, which is why it appears in the literature as $\log(|\kk|^2a^2)$. This is
the same division of labour as in \cite{Donoghue1994b}, where the coefficient of
$\log(-q^2)$ is a prediction of the massless sector while its scale is traded for the local
couplings. At finite
temperature in $d=3$ only the vanishing Matsubara frequency survives in the massless sector,
and it enters the free energy with weight $T$ instead of $\int d\omega/2\pi$, which is the
origin of the $(Ta)$ prefactor quoted in \cite{review} and is consistent with the
identification of the thermal zero mode as the source of the nonanalyticity in the
electromagnetic case \cite{FLMhighT}. Neither \cite{FLMdT} nor \cite{review} writes the
coefficient of that term. Since the $m=n=0$ Matsubara term of the finite temperature kernel of \cite{FLMdT} is the
$d-1$ dimensional zero temperature integrand with $T$ in place of $\int d\omega/2\pi$, the
present mechanism supplies it:
\begin{equation}
g^{(2)}_{\rm na}\big|_{d=3,\,T>0}=T\,a^{-4}\,G^{(2)}_{\rm N,na}\big(2,a|\kk|\big)
=-\frac{T}{16\pi a^{2}}\,|\kk|^{2}\log\big(|\kk|^{2}a^{2}\big)\,.
\label{eq:Tcoef}
\end{equation}

Finally, for $d=2$ at finite temperature one has $\Dpar=1$ and $C_1=0$, so that the leading
nonanalyticity is absent. At second order in the deformation this holds exactly: in one
dimension $\big[p\cdot(p+l)\big]^2=p^2(p+l)^2$, so that the two form factors in \eqref{eq:twoff}
coincide identically, $G^{(2)}_{\rm N}(1,l)=G^{(2)}_{\rm D}(1,l)$, and the infrared part of the
Neumann one in \eqref{eq:IRtwo} reduces to the scaleless Dirichlet integral. This is why the
$d=1$ entries of the Dirichlet and Neumann tables of \cite{FLMdT} coincide,
$b_2/b_0=\tfrac13+\pi^{-2}$ in both cases. The vanishing of $C_1$ in dimensional regularization
is the corresponding statement about \eqref{eq:decomp}, whose left hand side equals $1$ when
$D=1$, so that the massless bubble cancels against the remaining terms. Beyond second order,
where a massless propagator in one Euclidean dimension is infrared singular, the absence of
nonanalytic terms at $\Dpar=1$ would require a separate treatment.

To summarize, for $d=2$ at $T=0$ ($\Dpar=2$) the nonanalyticity is $|\kk|^2\log|\kk|^2$ and
the first order affected is $\partial^2$ \cite{FLMdT,FLMimperfect}; for $d=3$ at $T=0$
($\Dpar=3$) it is $|\kk|^3$, and the first order affected is $\partial^3$; for $d=3$ at $T>0$
($\Dpar=2$) it is $(Ta)|\kk|^2\log(|\kk|^2a^2)$, again at order $\partial^2$
\cite{FLMdT,review,FLMhighT}; and for $d=2$ at $T>0$ ($\Dpar=1$) there is no leading
nonanalyticity. The case $d=3$, $T=0$ is the one that had not been made explicit. It shows
how the gapless mode identified in \cite{FLMdT,review} acts at zero temperature in $3+1$
dimensions, and makes its effect quantitative: the term is subleading with respect to the
next-to-leading order, but it precedes the fourth order.

The infrared scale $\mu$ in \eqref{eq:D2} deserves a comment. In $\Dpar=2$ the divergence of
$B_2$ sits at $q\to0$, and is the familiar infrared behaviour of a massless scalar in two Euclidean dimensions. It is not a
formal blemish: the gap between two Neumann mirrors is a waveguide supporting a massless
propagating mode, and that mode correlates distant regions of the surfaces. The nonlocal
term is the statement that such correlations exist, and $\mu$ is set by whatever terminates
them, the lateral size of the mirrors or the crossover to the exterior region. Within a purely
local description $\mu$ cannot be determined, which is the sense in which the local DE has to
be completed by the zero mode. In
\cite{FLMimperfect} its role is played by the imperfection of the mirrors, which is why that
calculation is finite and diverges only in the perfect limit.

A related point concerns the strict zero of the zero mode. The $p=0$ mode of $\phi_0$, the
constant, is the orbit of the symmetry \eqref{eq:shift}, and by the remark after \eqref{eq:simil} it is an exact
zero mode of $-\partial^2+V$ for any $\psi$. Its Gaussian integral diverges and it must be
treated by a collective coordinate with the corresponding Faddeev-Popov factor, as the
translational zero mode is treated in soliton quantization; this is the meaning of the prime in
\eqref{eq:Gamma0}. Since it is a single degree of freedom, its removal affects the $\Dpar\ge3$
results only through terms of measure zero in \eqref{eq:Gamma2}, while in $\Dpar=2$ it is
entangled with the infrared logarithm above. That treatment concerns the constant mode only: the rest of $\phi_0$ is a whole massless field in the parallel directions, and its integration is the one performed above.

\section{Application to a cylinder in front of a plane}
\label{sec:examples}
Equation \eqref{eq:Gamma0} applies, as it stands, to any gap function: one computes
$V=\partial^2\sqrt\psi/\sqrt\psi$ and evaluates the determinant. We do this for a cylinder of
radius $R$, with its axis parallel to the plane and at a distance $L$ from it, in the regime
$L\ll R$ where the DE applies. The gap depends only on the coordinate $x_1$ across the axis,
and near the line of closest approach it is a parabola,
\begin{equation}
\psi=L+\frac{x_1^2}{2R}=L\,\big(1+\rho^2\big)\,,\qquad
\rho=\frac{x_1}{\ell}\,,\qquad \ell\equiv\sqrt{2LR}\,.
\label{eq:parabola}
\end{equation}
The length $\ell$ is the distance over which the gap doubles, and it is the only scale that
enters the zero mode sector. We write $\xi\equiv L/R$ throughout.

We first recall what the local expansion gives. The proximity force approximation,
$E_{\rm PFA}=-(\pi^2/1440)\int\psi^{-3}$, gives $E_{\rm PFA}=-(\pi^3/3840)\,\ell/L^3$ per
unit length. Its corrections have the form
$E=E_{\rm PFA}\,[1+\theta_1\xi+\theta_2\xi^2+\dots]$, and for Neumann conditions \cite{BEJK,Bordag2006,FLMdT}
\begin{equation}
\theta_1=\frac{7}{36}-\frac{40}{3\pi^2}=-1.16\,,
\label{eq:theta1}
\end{equation}
while the Dirichlet value is $\tfrac{7}{36}$. What matters here is
that the local expansion produces integer powers of $\xi$ only: a term with $2n$ derivatives is
$\int\psi^{-3}(\psi\partial)^{2n}$, and with \eqref{eq:parabola} each $\psi\partial$
contributes a factor $L/\ell=\sqrt{\xi/2}$.

The zero mode adds a term of a different kind. From \eqref{eq:parabola} and
\eqref{eq:simil}, with the Laplacian in one dimension, its potential is
\begin{equation}
V=\frac{1}{\ell^2}\,v(\rho)\,,\qquad
v(\rho)=\frac{1}{(1+\rho^2)^2}\,,
\label{eq:vcyl}
\end{equation}
which is positive, has range $\ell$, and decays as $\rho^{-4}$. Since $\ell$ is the only
scale in $V$, the zero mode determinant per unit time and per unit length of the cylinder is
$\mathcal E_0=\kappa/\ell^2$, with $\kappa$ a pure number, and therefore
\begin{equation}
\frac{\mathcal E_0}{E_{\rm PFA}}=\theta_{3/2}\,\xi^{3/2}\,,\qquad
\theta_{3/2}=-\frac{960\sqrt2}{\pi^3}\,\kappa\,.
\label{eq:theta32}
\end{equation}
Thus, before any computation, the zero mode is seen to contribute a half integer power of
$L/R$: its scale is the geometric mean of the two lengths of the problem, and no local term can
produce such a power. This is the counterpart, in this geometry, of the nonanalytic terms of an
effective field theory with a massless particle \cite{Donoghue1994a,Donoghue1994b}. The local
couplings, here the coefficients of the DE, produce the integer powers, while the loop of the
massless field produces a term that cannot be mistaken for any of them, with a coefficient
fixed by the low energy sector alone.

The sign of that coefficient is fixed as well. Removing the tadpole from
$\half\tr\log(1+G_0V)$ leaves $\half\tr[\log(1+G_0V)-G_0V]\le0$, since $-\partial^2+V$ is
positive by \eqref{eq:simil}. Hence $\kappa<0$ and $\theta_{3/2}>0$: the zero mode strengthens the
attraction, and the Neumann energy is
\begin{equation}
E=E_{\rm PFA}\Big[1+\theta_1\xi+\theta_{3/2}\,\xi^{3/2}+O(\xi^2)\Big]\,.
\label{eq:expansion}
\end{equation}
For perfect conductors the same expansion holds with $E_{\rm PFA}$ doubled, since the
electromagnetic problem separates exactly into the two scalar sectors \cite{EJKS2006} and the
Dirichlet one is analytic; hence $\theta_1^{\rm EM}=\tfrac{7}{36}-\tfrac{20}{3\pi^2}$ and
$\theta_{3/2}^{\rm EM}=\theta_{3/2}/2$.

It remains to compute $\kappa$. Since the potential depends on $x_1$ alone, the determinant
factorizes into one dimensional determinants on the line, one for each value of
$q^2=\omega^2+k_2^2$, where $k_2$ is the momentum along the axis, and each of them is given by
the Gel'fand-Yaglom formula \cite{Dunne2008}. At fixed $q$ the tadpole, that is, the scaleless
first order term of $\tr\log(1+G_0V)$ in \eqref{eq:expand}, equals $\frac{1}{2q}\int v=\frac{\pi}{4q}$. Removing it
exactly, and working in units $\ell=1$,
\begin{equation}
\kappa=\frac{1}{4\pi}\int_0^\infty\!q\,dq\,
\Big[\log\mathcal F(q)-\frac{\pi}{4q}\Big]\,,\qquad
\mathcal F(q)=\lim_{\rho\to\infty}\varphi(\rho)\,e^{-q\rho}\,,
\label{eq:kappacyl}
\end{equation}
where $\varphi$ is the solution of $-\varphi''+(v+q^2)\varphi=0$ that behaves as $e^{q\rho}$ for
$\rho\to-\infty$. The integral can be  evaluated by integrating the Riccati equation for the
logarithmic derivative of $\varphi$, and gives
\begin{equation}
\kappa=-0.02807\,,\qquad \theta_{3/2}=1.229\,.
\label{eq:numbers}
\end{equation}
We present more details in Appendix~\ref{app:kappa}.

Two features of this result deserve comment. First, $\kappa$ cannot be obtained from the second
order formula \eqref{eq:Gamma2}, that is, from the expansion in powers of $V$: here the second
order term is logarithmically divergent, because $\int v=\pi/2\neq0$ and a one dimensional
potential is never weak at long wavelengths. The parabola is not a small deformation of the
plane, and the determinant has to be evaluated as a whole. Second, the zero mode term is not
small: at $\xi=0.1$, $\theta_1\xi=-0.12$ while $\theta_{3/2}\xi^{3/2}=+0.04$, a third of the first
correction. It lies in the range covered by exact numerical calculations, and it should be
visible there.

A comparison with the numerical results of Ref.~\cite{LMV2008} requires some care. 
For the Neumann sector it was found there that the fitted linear coefficient depends 
strongly on the form assumed for the next correction and, in order to improve the fits, 
 a term $\xi^2\log\xi$ was included.  We have  compared the expansion above directly with the exact 
energies. 

To that end, we evaluated numerically the exact Neumann energy for the cylinder-plane geometry from the scattering formula of Ref.~\cite{LMV2008}, on a fine grid of values of $\xi$ between $0.005$ and $0.5$. The numerical error, estimated by varying the truncation of the partial-wave 
matrices, is below $10^{-8}$. As a check, this reproduces the linear fit quoted in their 
Fig.~12 on $0.04\le\xi\le0.07$ (a slope of $-0.784$, against their $-0.781$).

\begin{figure}[t]
\includegraphics[width=\columnwidth]{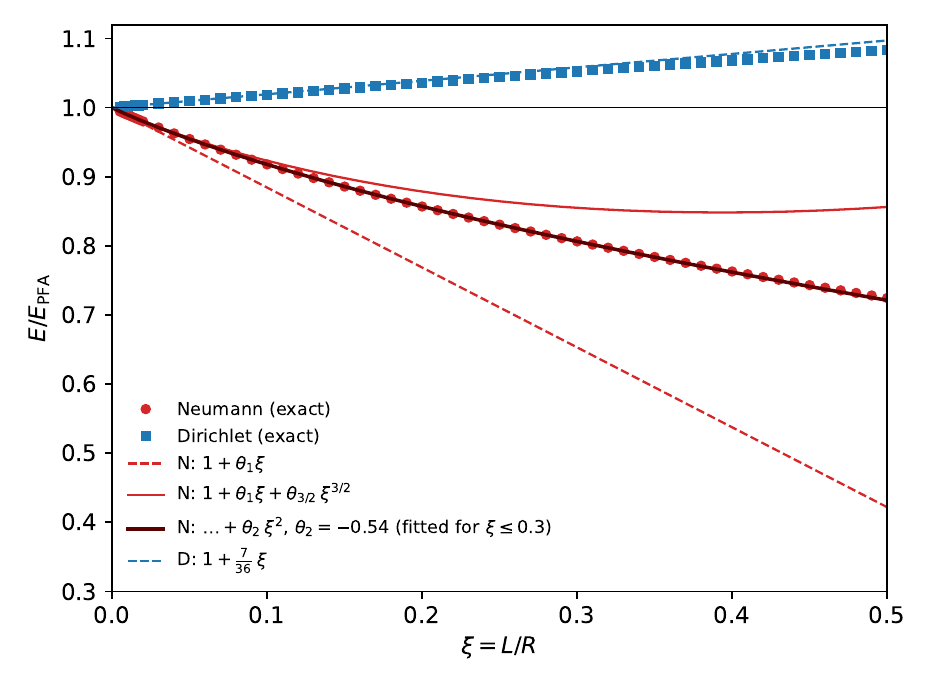}
\caption{Ratio $E/E_{\rm PFA}$ for a cylinder in front of a plane, for one scalar mode, as a function of
$\xi=L/R$. Points: exact energies for Neumann (circles) and Dirichlet (squares) conditions. Dashed lines:
$1+\theta_1\xi$ with the analytic $\theta_1$. Thin solid line: Neumann with the zero mode term,
$1+\theta_1\xi+\theta_{3/2}\xi^{3/2}$, with $\theta_{3/2}=1.229$. Thick solid line: the same with an
added $\theta_2\xi^2$, where $\theta_2$ is the only fitted parameter (fitted to the points with $\xi\le0.3$).}
\label{fig:energies}
\end{figure}

Figure~\ref{fig:energies} compares the exact energies with the successive approximations, for
Neumann and Dirichlet conditions. In the Neumann case, the linear approximation with the analytic
$\theta_1$ departs rapidly from the exact energy as $\xi$ grows. Adding the zero mode term
$\theta_{3/2}\,\xi^{3/2}$, with its analytic coefficient and no fitted parameter, improves the
agreement markedly in the range where the expansion can be expected to apply: at $\xi=0.2$, for
example, the error is reduced by about a factor of four. The remaining discrepancy has the sign and the
growth expected from the missing $\xi^2$ term. Including it, with $\theta_2$ as the only fitted parameter,
the curve follows the exact energies closely, to a fraction of a percent even beyond the range used
for the fit. The agreement is selective in the value of $\theta_{3/2}$: if it is displaced appreciably from
$1.229$ and $\theta_2$ is refitted, the curve no longer describes the data, and a fit that leaves both
coefficients free returns the analytic value of $\theta_{3/2}$. The Dirichlet energies, on the other hand,
follow the analytic expansion with no sign of a nonanalytic term.

The fitted $\theta_2$ should be regarded as an effective value, of order $-0.5$, and not as a
determination of the coefficient. Over the range used, the powers $\xi^2$ and $\xi^{5/2}$ are nearly
indistinguishable, so $\theta_2$ may absorb contributions of higher order, and its value depends on the
fitting window. A theoretical value would require the fourth order derivative expansion with the zero mode
kept explicitly. Regarding the logarithmic term introduced in Ref.~\cite{LMV2008} as a
fitting ansatz, our comparison suggests that it mimics the combination of the $\xi^{3/2}$ and $\xi^2$ terms
over a limited range. When both forms are given three free parameters, the one with $\xi^{3/2}$ describes the
exact energies better, recovers the analytic linear coefficient and extrapolates better to small $\xi$, whereas
the fits with the logarithm give a linear coefficient that differs appreciably from $\theta_1$. We do not
claim that a term of the type $\xi^2\log\xi$ is absent at higher orders, only that it is not needed to describe
these data.

\section{Discussion}
\label{sec:discussion}

Assembling the preceding sections, the procedure to compute the Casimir energy including the local  and nonlocal  contributions
is the following:

\begin{enumerate}
\item Flatten the gap with \eqref{eq:flatten} and expand in the Kaluza-Klein basis
\eqref{eq:basis}, separating $\phi_0$ from the massive levels.
\item Integrate the tower. It is gapped, $m_n\ge\pi/\psi$, so this step produces a local
expansion in $\psi\partial$. It yields the local part of the DE and the renormalization \eqref{eq:Z} of the zero mode kinetic operator,
$\psi\delta_{\mu\nu}\to Z_{\mu\nu}[\psi,\grad\psi]$, itself given by a derivative expansion. No
mass term for $\phi_0$ can appear, by the shift symmetry.
\item Integrate $\phi_0$ using \eqref{eq:simil} and \eqref{eq:Gamma0}, with $\psi$ replaced by
$Z$, and treat the constant mode by a collective coordinate.
\end{enumerate}

Step 3 requires a comment, since \eqref{eq:simil} was written for an isotropic
$Z_{\mu\nu}=\psi\delta_{\mu\nu}$. Writing $Z_{\mu\nu}=\psi(\delta_{\mu\nu}+z_{\mu\nu})$ with
$z_{\mu\nu}=O(\partial^2)$ by \eqref{eq:Z}, the isotropic part reproduces \eqref{eq:Gamma0} and
the anisotropy contributes to the nonanalytic part only at higher derivative order. We have not
constructed the generalization of \eqref{eq:simil} for anisotropic $Z$. The outcome has the
structure
\begin{equation}
F_{\rm Neumann}\;=\;\underbrace{F_{\rm loc}\big[\psi\big]}_{\text{local DE}}
\;+\;\underbrace{\half\log{\det}'\Big(-\partial^2+V[Z]\Big)}_{\text{nonlocal 
term}}\,,
\label{eq:master}
\end{equation}
so that the nonlocality is confined to a single massless determinant whose potential is a
local, DE computable functional of the geometry. At the order relevant for the leading nonanalytic terms, this closes the construction: with the
zero mode extracted, the rest is a conventional local expansion, and the extracted piece is a
standard object rather than an obstruction. This is the structure of any effective theory with
a massless particle \cite{Donoghue1994b,DonoghueElMenoufi}: local terms whose coefficients are
fixed by matching to the heavy sector, here the gapped tower, plus the loop of the light field,
whose couplings are themselves local.

For $\Dpar=2$ the remaining determinant could be treated exactly rather than perturbatively in
$V$, which would turn the $d=2$ case, where the zero mode enters already at next-to-leading
order, into a closed form answer
consisting of a local part plus a two dimensional Schr\"odinger determinant. That seems the
natural next calculation. 

The procedure also settles which boundary conditions admit a purely local fourth order DE.
From Section \ref{sec:KK} and the cases collected in Section \ref{sec:where},
Dirichlet-Dirichlet and Dirichlet-Neumann pairs are gapped, and their expansion is local. Neumann-Neumann is not, and
in $3+1$ dimensions at zero temperature the zero mode enters at order $\partial^3$, so a
fourth order construction applies to it, even at $T=0$, only with the zero mode kept
explicitly as in \eqref{eq:master}; Appendix \ref{app:check} verifies the $\partial^3$ term
against an independent calculation. The same applies to the polarization of a perfect conductor pair that reduces to
Neumann conditions, the one with reflection coefficient $+1$, whose $n=0$ mode is the TEM mode
of a parallel plate waveguide; the other polarization is Dirichlet-like and gapped. We label
the two sectors by their boundary condition throughout, since the transverse electric and
transverse magnetic labels are assigned oppositely in the Lifshitz convention and in
\cite{EHGK1,EHGK2}. Any classification of fourth order structures for the electromagnetic field
should therefore be stated for the Dirichlet polarization, or for materials with
dissipation, or at the price of carrying \eqref{eq:master} along.

Two remarks limit the scope of this conclusion. First, it concerns the gap 
function alone: with both mirrors flat, the only geometric datum is $\psi$, and 
$V$ is built from it. In the general
problem of two curved surfaces, configurations with the same gap may differ in the distribution
of curvature between the bodies, and the treatment of the zero mode then requires the two
surface version of \eqref{eq:Z}, which we have not considered here. Second, configurations with
$\grad^2\psi=0$, for which $V$ reduces to the slope term, receive a correspondingly smaller
contribution from the zero mode.

\section{Summary}

The nonlocal terms of the derivative expansion for Neumann boundary conditions are a
statement about the Kaluza-Klein spectrum of the gap. The DE is an expansion in $\psi\partial$,
and nonlocal terms of the kind studied here appear precisely when that spectrum is gapless.
Neumann has no gap, because the shift symmetry $\phi\to\phi+c$ is unbroken for an arbitrary
profile, which makes the massless mode symmetry protected rather than accidental. In the auxiliary field formulation the
same statement reads: one of the two eigenvalues of the surface kernel vanishes as $\psi p^2$ for Neumann, while it tends to a constant for Dirichlet, so that the one loop form factor contains a massless bubble in the first case and a scaleless integral in the
second.

The gapless sector is a massless field in the parallel directions, with the local gap as a
position dependent coefficient. Its kinetic term is fixed to second order by the Gaussian
integral over the massive modes, $Z_{ij}=\psi[\delta_{ij}-\tfrac13\partial_i\psi\partial_j\psi]$,
the coefficient $\tfrac13$ being $2\zeta(2)/\pi^2$. Through the identity
$-\partial_\mu \psi\partial_\mu=\sqrt \psi(-\partial^2+V)\sqrt \psi$, and with the geometric
measure $\sqrt g=\psi$ cancelling the accompanying contact term, its contribution is
$\half\log\det'(-\partial^2+V)$ with $V=\partial^2\sqrt\psi/\sqrt\psi$. The nonanalytic part
enters at derivative order $\Dpar$, with the coefficient \eqref{eq:CD}. This reproduces the
known cases, is consistent with the coincidence of the Dirichlet and Neumann next-to-leading order
coefficients at $\Dpar=1$, and supplies the missing case: a $|\kk|^3$ term at order
$\partial^3$, with the coefficient $-1/128$ of \eqref{eq:d3T0}, for Neumann in $3+1$ dimensions at zero temperature.

The prediction is verified in Appendix \ref{app:check}. In the form factor normalization of \cite{FLMdT} it reads
$G^{(2)}_{\rm N,na}(d,l)=-l^4B_d(l)/8$; for $d=2$ this reproduces the one nonanalytic
coefficient recorded in the literature, and for $d=3$ it agrees to eight figures with a direct
evaluation of the Neumann form factor. Independently, it is present with the predicted
coefficient $\pi^3/32$ in the second order corrugation kernel of \cite{EHGK1,EHGK2}, and absent
from its Dirichlet counterpart. A purely local fourth order derivative expansion is therefore
not available for Neumann conditions, or for the perfectly conducting polarization that reduces
to them, even at zero temperature. What is available is the effective description that keeps
the zero mode, which supplements the local expansion with terms that no local functional can
produce.

The picture is that of an effective field theory with a massless particle in the loop \cite{Donoghue1994a,Donoghue1994b}. The gapped Kaluza-Klein tower plays the
role of the heavy sector and yields the local DE; the zero mode is the light field, kept
explicitly, and its loop produces nonanalytic terms which are parameter free predictions of
the low energy theory. For a cylinder in front of a plane the prediction is a term
$\theta_{3/2}(L/R)^{3/2}$ in the ratio of the Neumann energy to the PFA, placed between the
first and second order terms of the local expansion, with $\theta_{3/2}=1.229$, and half that
value for perfect conductors. With that coefficient fixed, and only $\theta_2$ fitted, the
expansion describes the exact Neumann energies to a fraction of a percent up to $L/R=0.5$, and
it accounts for the logarithm that was postulated in \cite{LMV2008} to fit them.

\begin{acknowledgments}
This research was supported by Consejo Nacional de Investigaciones Cient\'ificas y
T\'ecnicas (CONICET).
\end{acknowledgments}

\appendix
\section{Consistency checks}
\label{app:check}
Of the consequences of the procedure, the cleanest one to test is 
\eqref{eq:d3T0}. For Neumann boundary conditions, 
the Casimir energy per unit area  for
a sinusoidally modulated gap,
$\delta\psi=\epsilon\cos(k x_1)$ at mean separation $\bar\psi$, should
contain, between the next-to-leading order $|\kk|^2$ term and the fourth order $|\kk|^4$ terms,
\begin{equation}
\frac{E_{\rm na}}{A}=-\frac{1}{256}\,\frac{\epsilon^2\,|k|^3}{\bar \psi^{2}}\,,
\label{eq:corrugpred}
\end{equation}
with no free parameter, the extra factor of $\half$ relative to \eqref{eq:d3T0} coming from the
mean square of the corrugation.

With the form factor normalization \eqref{eq:ffnorm} the statement to be 
tested is that of \eqref{eq:predff}, which the covariant treatment of Sections 
\ref{sec:exact} and \ref{sec:where} reproduces from \eqref{eq:rule}, as it 
must: for $\Dpar=2$ it gives $g^{(2)}_{\rm 
na}=-|\kk|^{2}\log(|\kk|^{2}a^{2})/16\pi a^{2}$, which is the result of
\cite{FLMdT} once the infrared scale is identified with $1/a$, and for $\Dpar=3$ it gives
$G^{(2)}_{\rm N,na}=-l^{3}/64$. The two routes agree, and the one nonanalytic coefficient
recorded in the literature comes out right. The $\Dpar=3$ coefficient is new, and we submit it
to two independent tests.

The first test uses the Neumann form factor itself. Equation \eqref{eq:twoff} at $d=3$
reduces to a two dimensional integral by using $q=|p+l|$ as a variable,
\begin{equation}
G^{(2)}_{\rm N}(3,l)=-\frac{1}{8\pi^{2}l}\int_0^\infty\!\!dp\;\frac{1}{1-e^{-2p}}
\int_{|p-l|}^{p+l}\!\!dq\;\frac{\big(p^{2}+q^{2}-l^{2}\big)^{2}}{e^{2q}-1}\,,
\end{equation}
which can be evaluated numerically. Two of its coefficients are known independently and serve
as controls: the value at $l=0$, $G^{(2)}_{\rm N}(3,0)=-\pi^{2}/120=12b_0$, with
$b_0=-\pi^2/1440$ the parallel plate coefficient, and the coefficient of $l^{2}$, which is
$2b_2$ with $b_2/b_0=\tfrac23-20/\pi^{2}$, the Neumann value given in \cite{FLMdT}
(where these coefficients are called $c_0$ and $c_2$). Subtracting both,
\begin{equation}
\lim_{l\to0}\;\frac{G^{(2)}_{\rm N}(3,l)-G^{(2)}_{\rm N}(3,0)-2b_2\,l^{2}}{l^{3}}
\;=\;-0.015625000\,,
\end{equation}
against the predicted $C_3=-1/64=-0.015625$.

The second test uses the second order corrugation kernel, and is independent of the formalism
used here and of \cite{FLMdT}. For two perfect 
mirrors, one flat and one corrugated with amplitude $\epsilon$ and wavelength 
$\lambda$ at mean separation $\bar\psi$, Refs. \cite{EHGK1,EHGK2} give the 
energy per unit area to second order in $\epsilon$ and for
all values of $s\equiv\bar\psi/\lambda$,
\begin{equation}
E_{\rm cf}=-\frac{\epsilon^{2}}{\bar \psi^{5}}\Big[G_{\rm D}(s)+G_{\rm N}(s)\Big]\,,
\end{equation}
in closed form in terms of polylogarithms of $u=e^{-4\pi s}$; $G_{\rm D}$ and $G_{\rm N}$ are
the two decoupled scalar sectors, labelled here by their boundary condition. With
$|\kk|=2\pi/\lambda$, matching normalizations gives
$G_{\rm D,N}(s)=-\tfrac14 G^{(2)}_{\rm D,N}(3,2\pi s)$, and the statements above become
statements about the small $s$ expansion of the two kernels. At order $s^0$ both must reduce to
the proximity force value, $G_{\rm D}(0)=G_{\rm N}(0)=\pi^{2}/480$, which is how \cite{EHGK1}
states its own limit. Neither sector may contain a term linear in $s$, and indeed, when fitted,
both coefficients are below $10^{-9}$. At order $s^{2}$ the $d=3$ entries of the two tables of
\cite{FLMdT}, $b_2/b_0=\tfrac23$ and $\tfrac23-20/\pi^2$, translate into
$\pi^{4}/1080=0.09019360$ for Dirichlet and $\pi^{4}/1080-\pi^{2}/36=-0.18396207$ for
Neumann, while the kernels give $0.09019365$ and $-0.18396151$. These three statements
validate the transcription and the normalization. The prediction is the fourth: at order
$s^{3}$ the coefficient must be $\pi^{3}/32=0.9689461$ for Neumann and zero for Dirichlet.
Subtracting the $s^{0}$ and $s^{2}$ pieces from the closed forms,
\begin{equation}
\begin{gathered}
\frac{G_{\rm N}(s)-\tfrac{\pi^{2}}{480}
-\big(\tfrac{\pi^4}{1080}-\tfrac{\pi^2}{36}\big)s^{2}}{s^{3}}
\;\overset{s=10^{-6}}{=}\;0.9689437\,,
\\[6pt]
\frac{G_{\rm D}(s)-\tfrac{\pi^{2}}{480}-\tfrac{\pi^{4}}{1080}\,s^{2}}{s^{3}}
\;\overset{s=10^{-6}}{=}\;-2.1\times10^{-7}\,,
\end{gathered}
\end{equation}
the Neumann residual converging to $\pi^{3}/32$ and the Dirichlet one vanishing linearly in
$s$, as it must if the Dirichlet tower is gapped. Continuing to the next order, both kernels
are analytic at $s^{4}$: the residuals after subtracting the $s^{3}$ terms tend to the
constants $-2.4650$ and $-0.20823$ with corrections linear in $s$, with no admixture of
$s^4\log s$ above the $10^{-6}$ level. The Dirichlet expansion is therefore analytic in
$|\kk|^2$ through order $|\kk|^4$, which is as far as the gap criterion has been tested.

The $|\kk|^{3}$ term of \eqref{eq:d3T0} is therefore present, with coefficient $-1/128$ and
no free parameter, and only for Neumann. It is contained in the closed form of \cite{EHGK1},
where it is the first correction beyond the $s^{2}$ term of the derivative expansion, and
appears to have gone unremarked. The nonlocal term predicted by the zero mode for Neumann
conditions at zero temperature in $3+1$ dimensions is thus not conjectural: it is a property
of a kernel that has been known in closed form for more than two decades, and the mechanism of
Sections \ref{sec:aux} to \ref{sec:where} accounts for both its origin and its magnitude, in
the way an effective theory with a massless particle fixes its nonanalytic terms from the low
energy sector alone.

\section{Evaluation of \texorpdfstring{$\kappa$}{kappa}}
\label{app:kappa}

We derive Eq.~(\ref{eq:kappacyl}) and describe its numerical evaluation. We work in
units $\ell=1$, so that $\psi=L(1+\rho^2)$ and the potential is
$v(\rho)=(1+\rho^2)^{-2}$.

As the potential depends only on $x_1$, 
the parallel momenta $(\omega,k_2)$ are conserved. Writing
$q^2=\omega^2+k_2^2$, the zero-mode energy per unit time and per unit length
of the cylinder is
\begin{equation}
E_0=\frac12\int\!\frac{d\omega\,dk_2}{(2\pi)^2}
\Big[\log\frac{\det(-\partial_\rho^2+q^2+v)}{\det(-\partial_\rho^2+q^2)}
-\mathrm{tadpole}\Big]
=\frac{1}{4\pi}\int_0^\infty\! q\,dq\,\Big[\log\mathcal F(q)-\frac{\pi}{4q}\Big],
\end{equation}
where $\int d^2q/(2\pi)^2=(2\pi)^{-1}\int q\,dq$ was used. The subtracted term
is the first-order term of $\tr\log(1+G_0V)$ in Eq.~(\ref{eq:expand}), $\mathrm{tr}(G_0V)
=\frac{1}{2q}\int v\,d\rho=\frac{\pi}{4q}$, which is scaleless in the full
problem and has been absorbed in the local expansion.

For the operator on the line, the ratio of
determinants is given by~\cite{Dunne2008}
\begin{equation}
\frac{\det(-\partial_\rho^2+q^2+v)}{\det(-\partial_\rho^2+q^2)}
=\mathcal F(q)=\lim_{\rho\to\infty}\varphi(\rho)\,e^{-q\rho},
\end{equation}
where $\varphi$ solves $-\varphi''+(v+q^2)\varphi=0$ with
$\varphi\sim e^{q\rho}$ as $\rho\to-\infty$. Since $q>0$, the operator
$-\partial_\rho^2+q^2+v$ is positive and has no zero mode.

With $y=\varphi'/\varphi$ one has
\begin{equation}
y'+y^2=v+q^2,\qquad y(-\infty)=q,
\qquad
\log\mathcal F(q)=\int_{-\infty}^{\infty}\!d\rho\,\big[y(\rho)-q\big].
\end{equation}
Using $\int v\,d\rho=\pi/2$, the subtraction can be done at the level of the
integrand,
\begin{equation}
\kappa=\frac{1}{4\pi}\int_0^\infty\! q\,dq\int_{-\infty}^{\infty}\!d\rho\,
\Big[y_q(\rho)-q-\frac{v(\rho)}{2q}\Big],
\end{equation}
which is what we integrate numerically: the Riccati equation is solved for
each $q$ with the initial condition $y=q+v/2q$ at a large negative $\rho$,
together with the running integral.

Since $v\sim\rho^{-4}$, the $\rho$ integral converges. For
$q\to\infty$ what remains after the subtraction is the second-order term,
$-\frac{1}{8q^3}\int v^2\,d\rho=-\frac{5\pi}{128\,q^3}$, so the integrand of
the $q$ integral decays as $q^{-2}$. For $q\to0$, $\log\mathcal F\sim-\log q$,
so the integrand of the $q$ integral tends to the constant $-\pi/4$. Both
ends are integrable, and the tails can be added analytically.
The result is
\begin{equation}
\kappa=-0.02807 ,
\end{equation}
from which $\theta_{3/2}=-960\sqrt2\,\kappa/\pi^3=1.229$ follows
by Eq.~(\ref{eq:theta32}).

\end{document}